\documentclass[twocolumn,nofootinbib]{revtex4-1}
\usepackage{latexsym,epsfig,amssymb, amsmath,nicefrac}

  \usepackage{amssymb}
  \usepackage{graphicx}
  \usepackage{amsmath}
  \usepackage{amsmath,amssymb,amsthm}
  \usepackage{verbatim}
  \usepackage{hyperref}

\def\bi{\begin{itemize}}
\def\ei{\end{itemize}}
\def\be{\begin{equation}}   
\def\ee{\end{equation}}
\def\ba{\begin{eqnarray}}   
\def\ea{\end{eqnarray}}

\def\lsim{\mathrel{\mathop
  {\hbox{\lower0.5ex\hbox{$\sim$}\kern-0.8em\lower-0.7ex\hbox{$<$}}}}}
\def\gsim{\mathrel{\mathop
  {\hbox{\lower0.5ex\hbox{$\sim$}\kern-0.8em\lower-0.7ex\hbox{$>$}}}}}

\newcommand{\cH}{\mathcal{H}}

\begin{document}

\title{{\Large The large-$N$ running of the spectral index of inflation}}

\author{ Juan Garcia-Bellido$^1$ and Diederik Roest$^2$}

\affiliation{{}$^1$ Instituto de F\'isica Te\'orica IFT-UAM/CSIC, Universidad Aut\'onoma de Madrid, Cantoblanco 28049 Madrid, Spain, juan.garciabellido@uam.es}

\affiliation{{}$^2$Centre for Theoretical Physics, University of Groningen, Nijenborgh 4, 9747 AG Groningen, The Netherlands, d.roest@rug.nl}

\begin{abstract}
We extend previous classifications of inflationary models by means of their behaviour at large-$N$, where $N$ is the number of e-foldings. In addition to the perturbative $1/N$ case, whose slow-roll parameters fall off as powers of $1/N$, we introduce the constant, non-perturbative and logarithmic classes. This covers the large majority of inflationary models. Furthermore, we calculate the running of the spectral tilt for all these classes. Remarkably, we find that the tilt's runnings essentially cluster around the permil level. We comment on the implications for future experiments.

\end{abstract}

\maketitle

\smallskip

\newpage

\section{Introduction}

After measuring the cosmic microwave background (CMB) with unprecedented precision, Planck has presented us with a pristine image of the early Universe. This has resulted in strong constraints on the inflationary phase. In particular, the spectral index has been measured to be \cite{Ade:2013}
 \begin{align}
  n_s & = 0.9603 \pm 0.0073 \,, \label{ns}
 \end{align}
although subsequent analyses have argued for a slightly higher value \cite{Spergel:2013}. Similarly, the running of the spectral index has been constrained with some precision:
 \begin{align}
 \alpha_s = - 0.0134 \pm 0.0090 \,.
 \end{align}
On the other hand, the BICEP2 collaboration~\cite{BICEP2} has recently announced the detection of the 
primordial gravitational wave background produced during inflation, with an amplitude responsible for a tensor-to-scalar ratio
\begin{align}
r = 0.20^{\,+0.07}_{\,-0.05} \,,
\end{align}
with $r=0$ ruled out at more than 5$\sigma$. This bound is in slight tension with the reported Planck upper bound
for a 6-parameter $\Lambda$CDM model, $r < 0.11$ at 95\% c.l., although foreground subtraction could decrease to some extent  the reported value. If confirmed, such detection would lead to important consequences. In particular,
many inflationary scenarios are ruled out by these bounds. Indeed, Planck and BICEP2 together seem to prefer large
field models of inflation~\cite{Martin:2013a, Martin:2013b}, with predictions different from the plateau potentials of Starobinsky-like models~\cite{Starobinsky:1982}, preferred before BICEP2 results.

In this paper we aim to understand the values quoted above in terms of the number of e-foldings $N$ between the horizon exit of the quantum modes that have resulted in the CMB anisotropies, and the end of inflation. As will be discussed in more detail below, this approach allows for a classification of different inflationary models in terms of universality classes with distinct large-$N$ behavior. It builds on preceeding works \cite{Mukhanov:2013, Roest:2013}, where the deviation from a Harrison-Zeldovich scale-invariant spectrum with $n_s = 1$ was attributed to $1/N$ effects. We will extend this with different leading terms in the $1/N$ expansion. Moreover, we will also investigate the running of the spectral index as a function of $N$. The running is the second-order term in the expansion of the scale-dependence of the power spectrum, and hence is natural to consider it in this framework as well. Finally, the values of this running is considered for a wide range of models, and shown to cluster in a very limited range around a central value of $\log_{10}|\alpha_s| = - 3.2$.

The outline of the paper is as follows. We introduce the $N$-formalism in section 2, where all slow-roll parameters are expressed in terms of $N$. Section 3 contains the different universality classes of large-$N$ behaviour. The running of the spectral index is discussed in section 4. Finally, we conclude in section 5. In the Appendix we consider also models of $k$-inflation as a separate class of its own.

\section{The $N$-formalism}

In this section we will express the time evolution of all cosmological observables through their $N$-dependence. In particular, we will demonstrate that it suffices to specify the equation of state parameter $\epsilon(N)$ to fully determine the inflationary phase. We will refer to this approach as the $N$-formalism. It can be seen as the background complement to the $\delta N$-formalism for cosmological perturbations \cite{Starobinsky:1982, Salopek:1990, Sasaki:1995, Lyth:2004}. Moreover, in the next section we will demonstrate its use in providing a perturbative expansion for inflationary models.

The evolution of a general scalar field is described by the Hamiltonian and momentum constraints
\be
3H^2(\phi) = 2 H'(\phi)^2 + V(\phi) \,, \hspace{0.5cm} \dot\phi = -2H'(\phi)\,, \label{Ham} 
\ee
plus the evolution equations
\be\label{epsilon}
\dot H = -2H'(\phi)^2\,, \hspace{0.5cm} \ddot\phi + 3H(\phi)\dot\phi + V'(\phi) = 0\,,
\ee
where dots denote derivatives w.r.t. cosmic time and primes w.r.t. the field $\phi$.
Throughout this paper we set $\kappa^2 = 8\pi G \equiv 1$. The only quantity needed to specify the whole evolution is therefore the Hubble function $H(\phi)$. The Hamilton-Jacobi formalism allows one to compute the evolution in terms of a new time-variable, the scalar field $\phi$.
In general, these equations are too difficult to solve without specifying a potential. However, we can characterize the whole evolution via the equation of state parameter
\be
\epsilon = - \frac{\dot H}{H^2} =  \frac{3}{2}(1+w)\,,
\ee
and its derivatives, the so called Hubble slow-roll parameters,
\ba\label{srparam}
\epsilon &=& 2\left(\frac{H'(\phi)}{H(\phi)}\right)^2 = \frac{\dot\phi^2}{2H^2}  \, \notag \\
\delta &=& 2\left(\frac{H''(\phi)}{H(\phi)}\right) = - \frac{\ddot\phi}{H\dot\phi}  \, \notag \\
\xi &=&  4\left(\frac{H'(\phi)H'''(\phi}{H^2(\phi)}\right) = \frac{\stackrel{\dots}{\phi}}{H^2\dot\phi} - \delta^2  \,,
\ea
In the slow-roll approximation, one is assuming $H^2(\phi) \propto V(\phi)$. Therefore the Hubble slow-roll parameters are related to the slow-roll parameters $\epsilon_V, \eta_V, \ldots$ defined in terms of derivatives of the potential by replacing $H^2(\phi)$  by $V(\phi)$ in the above. 

The two sets of slow-roll parameters are equivalent in the slow-roll approximation. However, the Hubble slow-roll parameters offer a number of advantages for our purposes. First of all, they are more accurate since they do not ignore the scalar kinetic term in \eqref{Ham}. Moreover, they can be derived from the equation of state parameter,
\ba
H(N) &=& H_0\,\exp\,\int\epsilon(N) dN \,, \notag \\
\delta(N) &=& \epsilon(N) + \frac{1}{2}\frac{\epsilon'}{\epsilon}(N)\,, \notag \\
\xi(N) &=& \frac{3}{2}\epsilon'(N) + \epsilon^2(N) + \frac{1}{2}\left(\frac{\epsilon'}{\epsilon}\right)'(N)\,, 
\ea
where primes denote differentiation w.r.t $N$, the number of $e$-folds
\be
N = \ln\frac{a_{\rm end}}{a_{\rm CMB}} =  \int_{t_{\rm CMB}}^{t_{\rm end}} H dt= \int_{\phi_{\rm CMB}}^{\phi_{\rm end}} \frac{d\phi}{\sqrt{2\epsilon}}\,.
\ee
We can thus describe the whole evolution in a new time unit, related to the scale factor, and not directly to cosmic time. The field $\phi$ therefore no longer plays the role of the clock; instead time is measured by the logarithmic growth of the scale factor, see e.g.~also \cite{stochastic}. We can also write the conformal time $\tau$ as an integral 
\ba\label{conftime}
-\tau\cH & = & 1 + e^{-\int dN(1-\epsilon)}\int dN \epsilon(N)e^{\int dN(1-\epsilon)}  \notag \\[2mm]
&\!\simeq\!& \frac{1}{1-\epsilon(N)} \,,
\ea
where $\cH=aH$ and the last expression is valid only in the slow-roll approximation, where $\epsilon \ll 1$.

Note therefore that the equation of state parameter $\epsilon(N)$ is all we need to specify the dynamics.
In particular, this parameter determines the spectra of metric perturbations, both scalar and tensor,
\ba
P_s(k) & = &  \frac{H^2}{8\pi^2\epsilon}\,,\qquad
P_t(k) = \frac{2H^2}{\pi^2}\,.
\ea
The scale dependence of the power spectra can be expanded as
\ba
 \ln P_s(k) & = & \ln P_s(k_0) + (n_s - 1) \ln \frac{k}{k_0} + \tfrac12 \alpha_s \ln^2 \frac{k}{k_0} \,, \notag \\
 \ln P_t(k) & = & \ln P_s(k_0) + n_t \ln \frac{k}{k_0} + \tfrac12 \alpha_t \ln^2 \frac{k}{k_0} \,.
  \ea
This translates into the following definitions for the scale-dependence coefficients in terms of $\epsilon(N)$:
\ba
&& n_s - 1 = - 2\,\epsilon(N) + \frac{\epsilon'}{\epsilon}(N) \,, \hspace{1cm} n_t  = - 2\,\epsilon(N)\,, \notag \\
&& \alpha_s =  2\epsilon' - \left(\frac{\epsilon'}{\epsilon}\right)'  \,, \hspace{2.3cm} \alpha_t = 2\epsilon'\,.
 \ea
which can  be computed straightforwardly in the $N$-formalism. Note that the $only$ difference between the scalar and tensor spectral indices is the term $\epsilon'/\epsilon$. In models where $\epsilon$ is constant in $N$, they are identical, and this can be used to constrain the amplitude of the tensor contribution, while in general $\epsilon'/\epsilon$ is not constant. Finally,
\ba
r & = & 16\,\epsilon(N)  
\ea
denotes  the ratio between the scalar and tensor power spectra.

\section{Universality classes}

In this section we will discuss the $N$-dependence of the equation of state parameter $\epsilon$ of a number of inflationary models. In particular, we will classify such models according to the leading behaviour at large values of $N$. Although non-trivial to determine observationally, typical values for $N$ range around $50$ and $60$. Furthermore, this number does not constitute an upper bound on $N$, but only corresponds to the portion of the inflationary trajectory that we have observational access to at the moment via the CMB. The total number of e-foldings could thus be much larger. Barring the apparent hints of power loss at large angular scales we have no reason to assume that $N=50-60$ is by any means special and hence it seems reasonable to assume that inflation has taken place over more than the observed number of e-foldings.

Moreover, the recently measured value of the tensor-to-scalar ratio by BICEP2 collaboration, $r\sim0.2$, suggests that inflation took place at high energy scales, close to $10^{16}$ GeV, and thus the number of e-folds of inflation has a robust lower limit of $N\gsim50$, which generically indicates large $N$ values for cosmological quantities that crossed the Hubble scale during inflation and reenter today. It therefore seems natural to consider a perturbative expansion of inflationary observables in terms of $1/N$. In the case of a polynomial expansion, as was analyzed in Refs.~\cite{Mukhanov:2013, Roest:2013}, one can argue that only the leading contributions are relevant.\footnote{In particular, this class was analyzed in terms of the Hubble slow-roll parameters by \cite{Mukhanov:2013}, while \cite{Roest:2013} employs the slow-roll parameters in terms of the scalar potential. In this paper we will follow the former approach.} Subleading corrections will be very difficult to measure observationally. Moreover, strictly within the slow-roll approximation, these are not unambiguously defined for the following reasons.\footnote{We thank Andrei Linde for an interesting discussion on this issue.} First of all, the approximation itself is precisely obtained by neglecting this type of subleading terms and equating the Hubble parameter and the scalar potential. Moreover, $N$ can only be defined up to order one errors even within the slow-roll approximation. Therefore any subleading coefficients of a polynomial expansion of the slow-roll result are physically meaningless; these can only be obtained by performing a full Hamilton-Jacobi analysis.

The large-$N$ limit therefore seems to be a powerful discriminant between different models. We will introduce a number of new classes of models, where each class has a specific $1/N$-dependence. Physically different models in the same class will thus give rise to the same large-$N$ behaviour, and generically agree on their cosmological predictions\footnote{See \cite{Kiritsis} for a similar classification from a holographic viewpoint.}. The reason for this universality between different models is that the large-$N$ limit only probes a limited region of the inflationary potential; for instance, for models where inflation takes place on a plateau, the details of the valley are washed out in this limit. Similarly, for chaotic-like inflationary models with polynomial potentials, the large-$N$ limit only depends on the highest power in the inflaton field.

Our discussion extends the aforementioned classification where only a leading $1/N^p$ behaviour for $\epsilon$ was considered, which we will refer to as the {\it perturbative} class. In addition to this possibility, we also introduce the {\it constant}, {\it non-perturbative} and {\it logarithmic} classes. For each of these, we will give the (leading order approximation to) the spectral tilt, tensor to scalar ratio and running. Moreover, for each case examples of inflationary models and (when possible) their exact  expressions are provided. We also assess the accuracy of the leading $1/N$ approximation.

\subsection{Constant class}

The constant class is characterized by a {\it constant}, $N$-independent equation of state parameter
\be
\epsilon(N) = \epsilon_0 \,.
\ee
The spectral index, tensor ratio and scalar running are given by
\be
n_s = 1 -2 \epsilon_0 \,, \hspace{5mm} r = 16\,\epsilon_0 \,,  \hspace{5mm} \alpha_s = 0\,.
\ee

An example of this class is power-law inflation, with an exponential scalar potential:
 \begin{align}
  V = V_0 e^{\alpha \phi} \,.
\end{align}
Although this model does not have a scenario for inflation to end by itself, one can study its predictions during the inflationary phase. These give rise to an exactly constant equation of state with $\epsilon_0 = \tfrac12 \alpha^2$. 
This class is ruled out since fixing the constant to agree with the scalar spectral index $n_s = 0.96$ one finds that $r=0.32$, which exceeds the upper limit set by Planck. This is illustrated by the upper line in Fig.~1.

\subsection{Perturbative class}

This class is characterized by a {\em perturbative} equation of state parameter
 \be
  \epsilon(N) = \frac{\epsilon_p}{N^p} \,,
 \ee
with $\epsilon_p$ constant. 
We will assume that $p\geq1$ as we do not know any viable model with $p<1$. This leads to the following observables for $p=1$:
\begin{align}
   n_s = 1 - \frac{2 \epsilon_1+1}{N} \,, \quad
   r = \frac{16 \epsilon_1}{N} \,, \quad  
  \alpha_s = - \frac{2 \epsilon_1+1}{N^2} \,. \label{perturbative}
 \end{align}
while $p>1$ gives rise to a qualitatively different behaviour:
 \begin{align}
   n_s = 1 - \frac{p}{N} \,,
\quad
   r = \frac{16 \epsilon_p}{N^p} \,, \quad  
  \alpha_s =  - \frac{p}{N^2} \,.
 \end{align}
An attractive feature of these universality classes is that the $1/N$ term provides a natural explanation for the percent deviation from scale invariance ($n_s \simeq 0.96$). The coefficients in this expansion, which are $\epsilon_1$ or $p$ respectively, are therefore order one, giving a natural perturbative expansion. Examples of corresponding inflationary models are:

\bigskip

\noindent
$\bullet$ {$p=1$}: \\
Chaotic inflation \cite{chaotic} with $$V = M^4 \left( \frac{\phi}{\mu} \right)^n$$ has an equation of state parameter given by
\ba
\epsilon = \frac{\tfrac14n}{N + \tfrac14 n} \,,
\ea
which is exact in the slow-roll approximation (with inflation ending at $\epsilon=1$).
This set of models is almost ruled out since fixing the constant to agree with the scalar spectral index $n_s = 0.96$ one finds that the models are just outside the 2$\sigma$ contour allowed by Planck, see the second highest line in Fig.~3.

\begin{figure}[t!]
\begin{center}
\includegraphics[width=8cm]{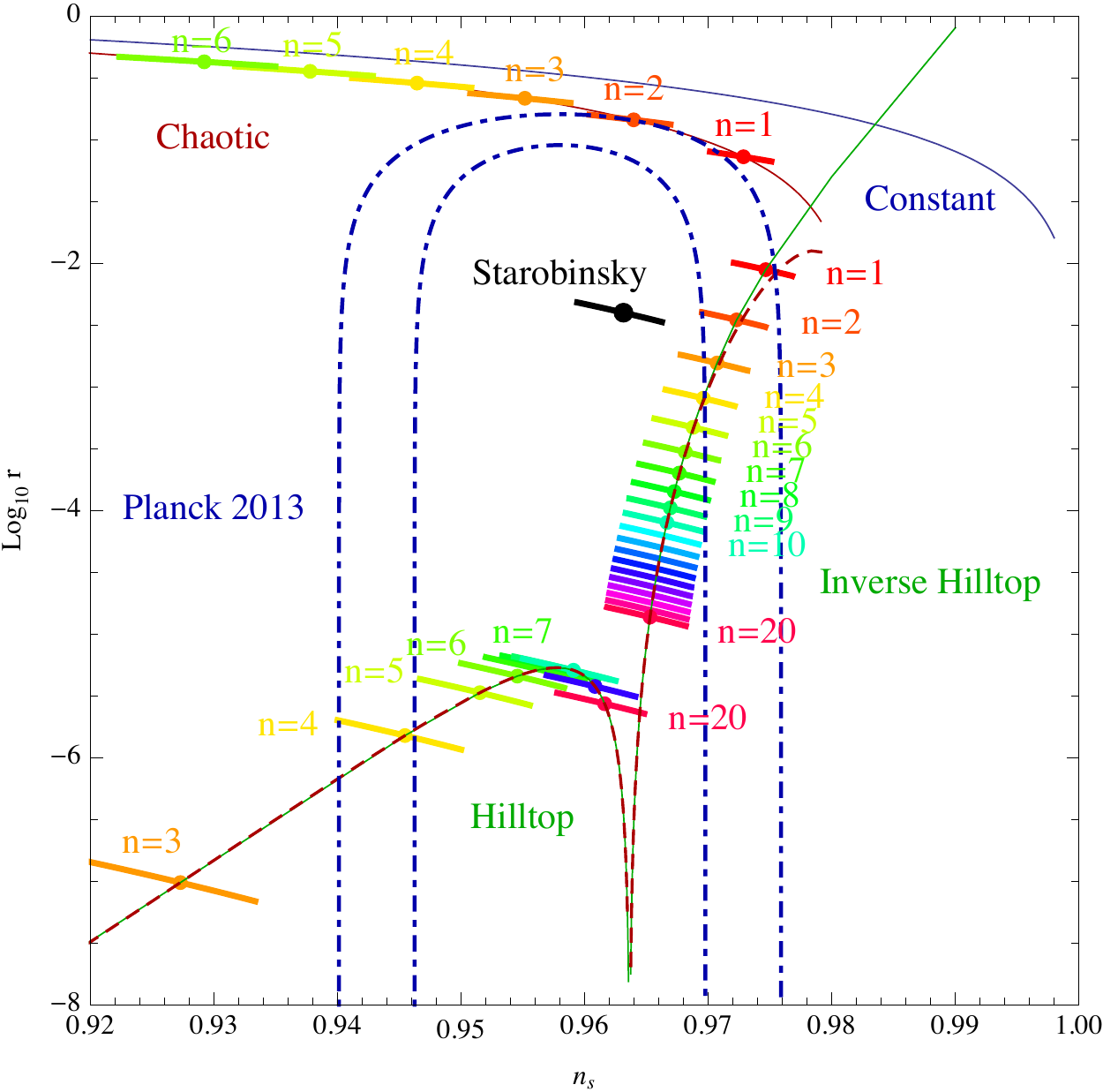}
\caption{The plane $(n_s, \log_{10} r)$ with the different perturbative models: chaotic, inverse hilltop and hilltop (both with $\mu=1$), as well as the constant model. The colored bars correspond to the range of $N\in [50,60]$. The solid lines are exact while the dashed lines are the leading-$N$ contributions. Note that these essentially agree inside the Planck 2013 region. Also shown for reference is the model of Starobinsky.}
\end{center}
\label{fig.nsr}
\end{figure}

\bigskip

\noindent
$ \bullet$ {$1<p<2$}: \\
Another type of models, that we will refer to as inverse hilltop, is characterized by a potential
 \begin{align} 
  V = V_0 \left[1 - \left( \frac{\mu}{\phi} \right)^n\right] \,, \label{hilltop-V}
 \end{align}
where $n$ is a positive power. It leads to an equation of state \cite{Martin:2013a}
 \begin{align}
\epsilon(N) = \frac{\tfrac12 n^2 / \mu^2}{\Big(Nn(n+2) / \mu^2\Big)^{\frac{2(n+1)}{n+2}}}  \,, \label{hilltop-E}
 \end{align}
at lowest order in $1/N$. Different positive values of $n$ interpolate from $p=1$ to $p=2$. 
Note that this model, for $n \geq4$, gets a spectral index close to $n_s=0.96$, well within the 2$\sigma$ contour of Planck, see the lower line in Fig.~2.

\bigskip

\noindent
$\bullet$ {$p=2$}: \\
The Whitt potential corresponding to the Starobinsky model \cite{Whitt:1984, Starobinsky:1982}
 \begin{align}
  V = V_0 \left(1 - e^{-\sqrt{2/3} \phi}\right)^2 \,,
 \end{align}
is characterized by an equation of state parameter
\be
\epsilon(N) = \frac{3}{4N^2}\,,
\ee
at lowest order in $1/N$. This model is not ruled out but precisely agrees with the scalar spectral index $n_s = 0.96$, and one finds that the model is well inside the 2$\sigma$ contour allowed by Planck.

An interesting generalisation in the same universality class is provided by T-models of inflation with a potential \cite{Kallosh:2013hoa}
$$V = V_0\tanh^{2n}(\phi/\sqrt{6\alpha})\,,$$ with $\alpha$ arbitrary, which corresponds to an equation of state parameter
\be
\epsilon(N) = \frac{3\alpha n}{4n N^2+2N\sqrt{3\alpha(3\alpha+4n^2)}+3\alpha n}\,.
\ee
This is an exact expression within the slow-roll approximation, with inflation ending at $\epsilon = 1$.
Note that this model gets a spectral index $n_s=0.96$ for $n=2$, almost independent of $\alpha$. As we vary the parameter $\alpha$ from 0.1 to 10, we move up in the plane $(n_s,\,r)$, as can be seen by the dot-dashed lines in Fig.~3.

\bigskip

\noindent
$\bullet$ {$p>2$}: \\
In contrast to the previous models, hilltop inflation takes place near the origin \cite{Boubekeur:2005}. Its scalar potential is 
 \begin{align} 
  V = V_0 \left[1 - \left( \frac{\phi}{\mu} \right)^n\right] \,, \label{hilltop-V}
 \end{align}
with $n>2$. It leads to an equation of state
 \begin{align}
  \epsilon(N) = \frac{\tfrac12 n^2 / \mu^2}{\Big(Nn(n-2) / \mu^2\Big)^{\frac{2n-2}{n-2}}} \,. \label{hilltop-E}
 \end{align}
Different values of $n$ therefore fill out the range $p>2$. For $n \geq 4$ this model is well within the Planck 2$\sigma$ contour.

\bigskip

We have illustrated the accuracy of the $1/N$ expansion in Fig.~2, where we plot the various perturbative models (Chaotic, Hilltop and Inverse hilltop inflation) as a function of their parameter (the leading power of the potential) in the plane ($n_s, \log_{10} r$). It is clear that, for models within the Planck-2013 2$\sigma$ region, the leading $1/N$ contribution describes correctly the model.

\subsection{Non-perturbative class}

This class is characterized by an equation of state parameter that asymptotes to
\be
\epsilon(N) = \epsilon_0\,e^{-2c N}\,,
\ee
which is non-perturbative around $1/N\to0$.
The resulting spectral index, tensor ratio and scalar running at lowest order are given by
\begin{align}
n_s  = 1 -2c \,, \quad 
\alpha_s = -4c\,\epsilon_0\,e^{-2c N}\,, \quad
r = 16\,\epsilon_0\,e^{-2c N} \,. \label{non-perturbative}
\end{align}
In contrast to the previous class, the non-perturbative class has a constant shift of the spectral index. The corresponding parameter thus has to be percent level to account for the observed deviation from spectral index. This spells somewhat of a problem of perturbation theory, as the effective parameter in which we are expanding $\epsilon$ is not $1/N$ but $1/(cN)$. While the former is naturally small, the latter is not necessarily. We therefore expect subleading terms to have some relevance in this region of parameter space. Indeed this will be confirmed by an analysis of the following two examples.

For the new inflation model \cite{Linde:1981mu} with
 \begin{align}
  V = V_0 \left( 1- \frac{\phi^2}{\mu^2}\right)^2 \,,
\end{align} 
where $c=4/\mu^2$ and $\epsilon_0=2c(1-\sqrt{c})^2$, the above expression for $\epsilon$ is exact and not a large-$N$ approximation. This model is almost ruled out, since fixing the constant $c$ to agree with the scalar spectral index $n_s = 0.96$ one finds that the models are just inside the 2$\sigma$ contour allowed by Planck, see Fig.~2.

\begin{figure}[htb]
\begin{center}
\includegraphics[width=8cm]{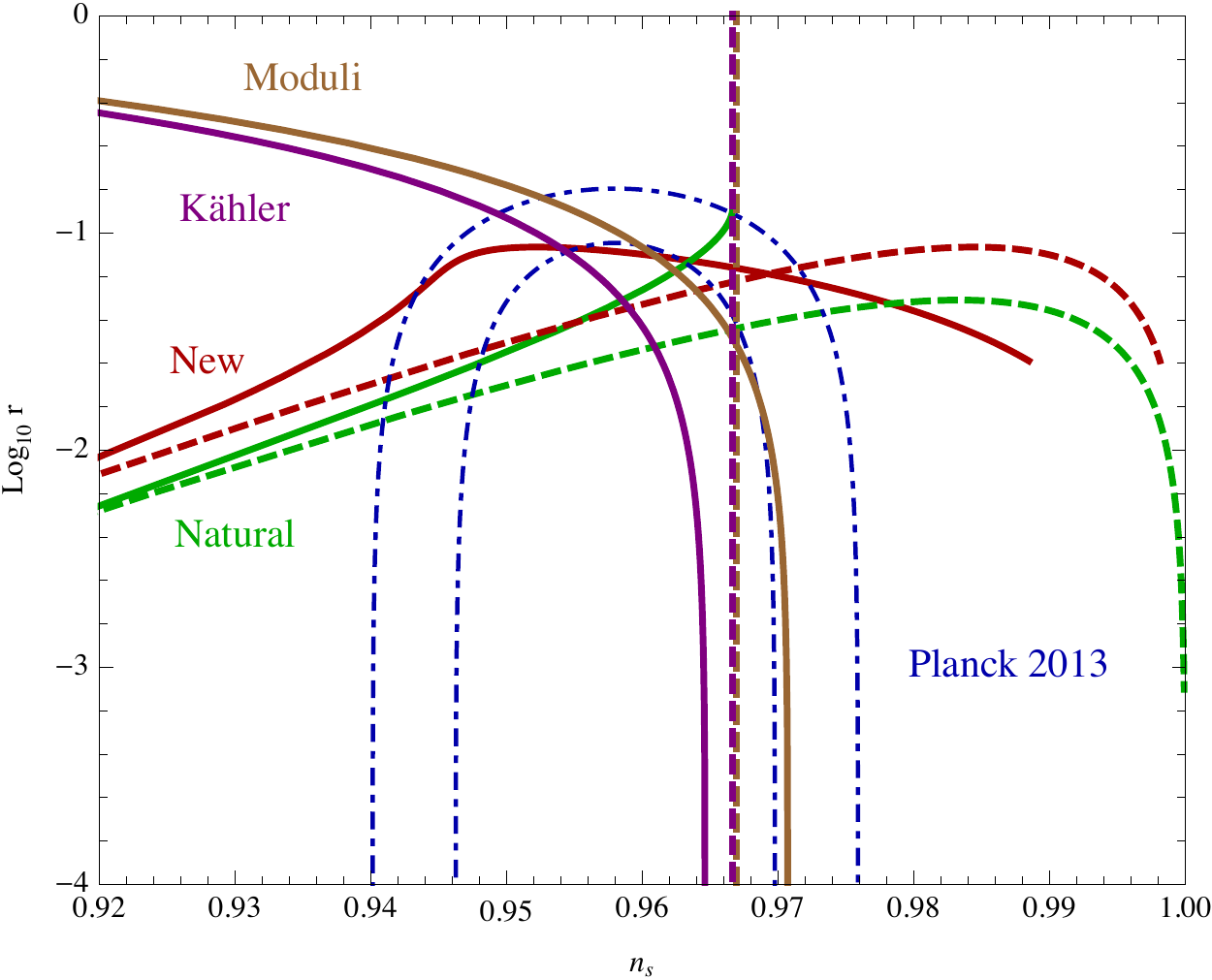}
\caption{The plane $(n_s, \log_{10} r)$ with the different non-perturbative models: new and natural, as well as the logarithmic models: Moduli and K\"ahler. The solid lines correspond to the exact expressions, while the dashed lines are the leading $1/N$ contributions. Note that these differ significantly even inside the Planck 2013 region.}
\end{center}
\label{fig.nsr2}
\end{figure}

Another example of this class is natural inflation \cite{natural} with
 \be
  V =  V_0 \left(1 + \cos\frac{\phi}{\mu}\right) \,.
 \ee
The equation of state parameter can be computed exactly as
\ba
\epsilon & = & \frac{c}{e^{2c N}-1}\,,
\ea
with the $c=1/ (2 \mu^2)$. 
This model is on the verge of being ruled out, since fixing the constant to agree with the scalar spectral index $n_s = 0.96$, one finds that the model is just inside the 2$\sigma$ contour allowed by Planck, see Fig.~2. It is different from that of New inflation since $\alpha_s = - c\,r/4$ is not satisfied in this case $\forall N$, since $(\epsilon'/\epsilon)'\neq0.$

We show in Fig.~2 the validity of the $1/N$ expansion for the various non-perturbative models (New and Natural inflation) as a function of their parameter (the v.e.v. of the field) in the plane ($n_s, \log_{10} r$). It is clear that, for models within the Planck-2013 2$\sigma$ region, the leading $1/N$ contribution gives a relatively poor description of the model dependence. In these models, the full non-perturbative expression is needed, and we cannot rely in the leading $1/N$ contribution.

\subsection{Logarithmic class}

This class is characterized by an equation of state parameter with logarithmic terms
\be
\epsilon(N) = \epsilon_p \frac{\ln^q\!N}{N^p} \,.
\ee
We will mainly be interested in the case with $p=2$. Then the leading expressions for the cosmological observables are
 \begin{align}
  n_s = 1 - \frac2N \,, \quad \alpha_s = - \frac{2}{N^2} \,, \quad 
  r = 16 \epsilon_p  \frac{\ln^q\!N}{N^2} \,.
 \end{align}
Remarkably, the logarithmic dependence drops out of the scalar power spectrum in the large-$N$ limit. The only remnant can be found in the ratio of the tensor to scalar power.

K\"ahler moduli inflation~\cite{Kahler} class, in particular, is characterized by a potential
\be
V = V_0\left(1 - \alpha \phi^{4/3}\,e^{-\beta\phi^{4/3}}\right)\,,
\ee
which corresponds to an equation of state parameter
\be
\epsilon(N) = \frac{c}{2N^2\sqrt{\ln\!N}} \,,
\ee
with $c=9/16\beta^{3/2}$. 
A toy version of this model has the scalar potential \cite{Martin:2013a}
 \begin{align}
  V = V_0 ( 1 - \alpha \phi e^{-\phi}) \,.
 \end{align}
Its inflationary trajectory can be characterized by the equation of state parameter
\be
\epsilon(N) = \frac{\ln^2\!N}{2N^2} \,.
\ee
 Note that both models  predict a spectral index $n_s=0.96$, and negligible tensor-to-scalar ratio, much smaller even than Starobinsky model, well within the 2$\sigma$ contour of Planck, see Fig.~2.

\section{Running of the spectral tilt}

In this section we will take a somewhat more phenomenological approach to study the possible values of the running $\alpha_s$ of the spectral index, independent of the $N$-formalism and universality classes. In the preceding sections we studied the parameter dependence of the $(n_s,r)$ values for different models. We will now fix these parameters to their best values to agree with the Planck and BICEP2 data on $(n_s,r)$, and subsequently calculate the resulting running $\alpha_s$ within that model. 

As we have given all relevant expressions in the previous section, we will only quote the results here. Firstly, in Fig.~3a the spectral index and the tensor-to-scalar ratio are plotted for all models that we have discussed. In case these models have free parameters we have fixed these in order to comply with the Planck data. Moreover, we have varied $N$ from $50$ to $60$ to give an indication of the $N$-dependence of these models. Secondly, in Fig.~3b we show the spectral index and its running, for the same parameter values.

\begin{figure}[t!]
\begin{center}
\includegraphics[width=8.5cm]{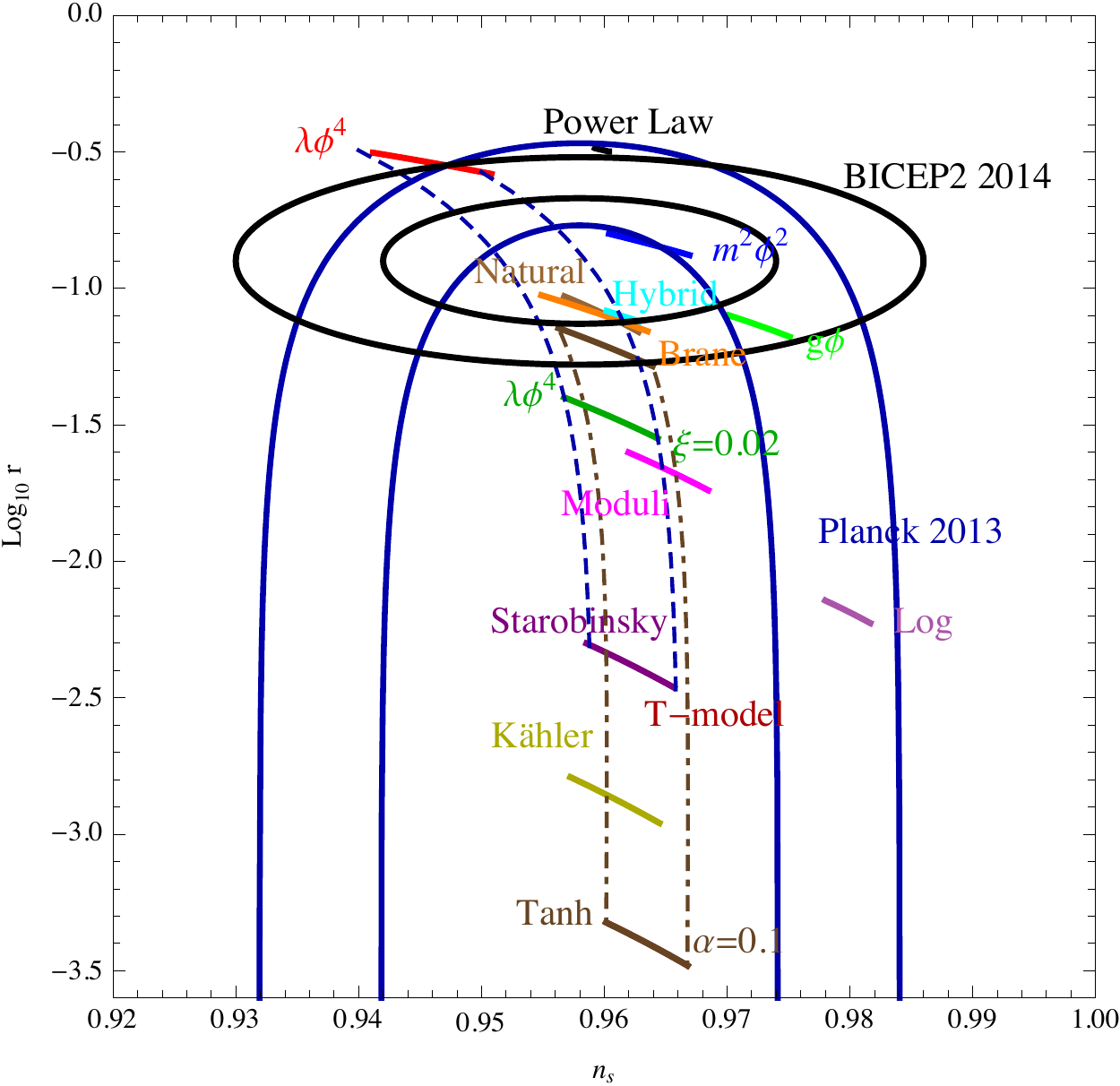}
\includegraphics[width=8.5cm]{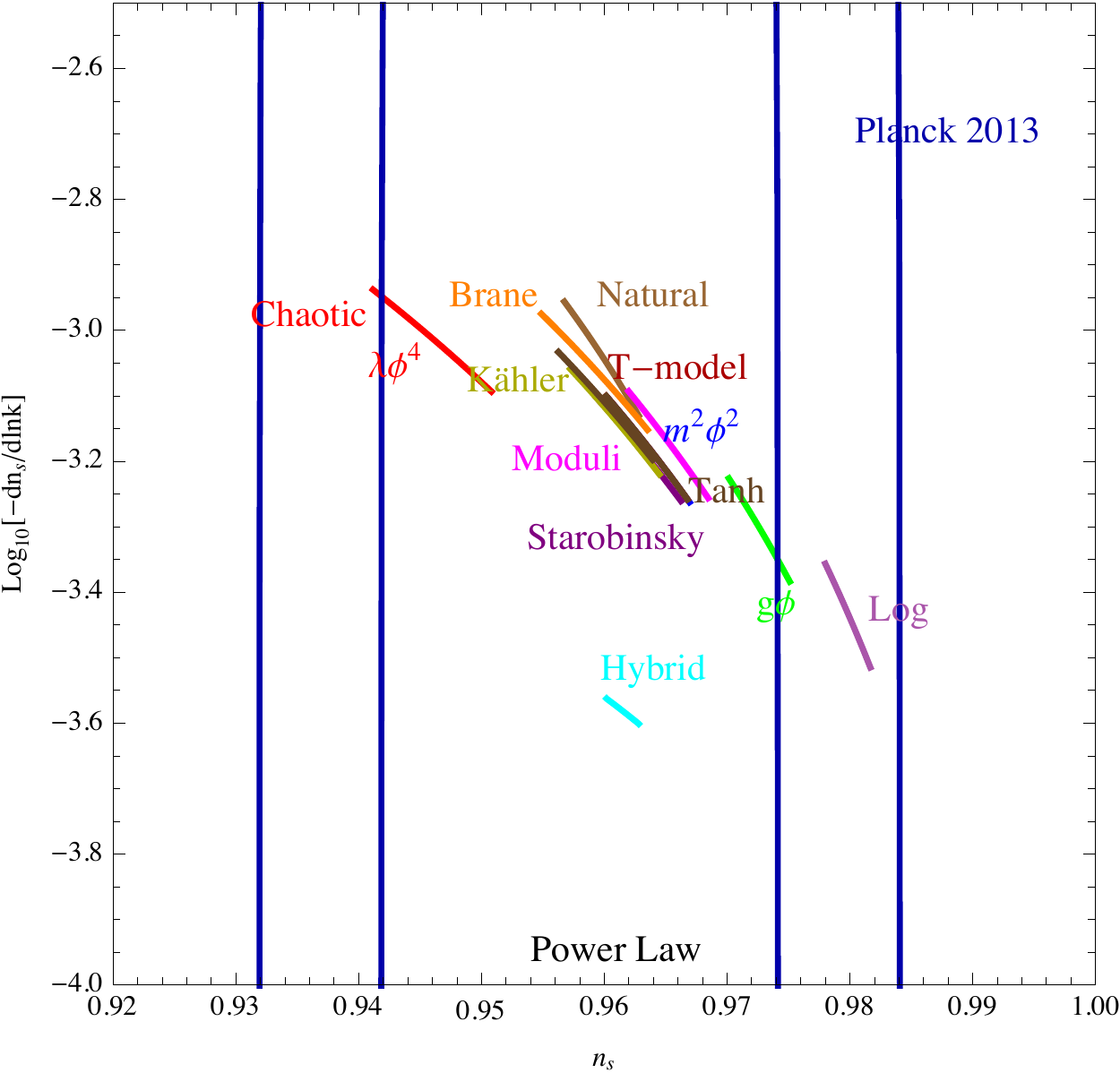}
\caption{The planes $(n_s, \log_{10} r)$ and  $(n_s, \log_{10}(-\alpha_s)$ with all the models discussed in the text. The ranges of values correspond to the interval $N\in[50,60]$. The solid blue (black) lines indicate the Planck (BICEP2)constraints at one and two sigma. The dashed line corresponds to non-minimally coupled chaotic $\lambda\phi^4$ model, $\xi=0 \to 10$. The dot-dashed line corresponds to T-models in a range of values of $\alpha=0.1 \to 10$.}
\end{center}
\label{fig.nsr}
\vspace{-0.5cm}
\end{figure}

While the ratio of tensor to scalar amplitudes in all models discussed above, which fall in various universality classes, can vary significantly over several orders of magnitude, see Fig.~3a, it is surprising that the actual running of the scalar spectral index $n_s$ does not, as shown in Fig.~3b. All models seem to cluster around $$\alpha_s \simeq - 0.001\,.$$ PRISM~\cite{PRISM1, PRISM2} is possibly capable of detecting the running of such models,  but unfortunately Planck has not enough resolution.

One could try to understand this clustering from the fact that we are only considering models that fall within the Planck bounds for ranges of e-folds from 50 to 60, i.e. $\Delta N \sim 10$, while the variation of the spectral index in all these models is never larger than their deviation from Harrison-Zel'dovich, i.e. $\Delta n_s \sim 0.01$. Therefore, it is not surprising that all of these models seem to fall in the same ballpark, irrespective of which universality class they belong to. This is a purely phenomenological observation. For instance, if we had considered sugra hybrid inflation, which has a large variation of the spectral index over CMB scales, from $n_s>1$ on horizon scales down to $n_s<0.8$ on small scales, then we would have predicted a $dn_s/d\ln k \sim 0.1$, orders of magnitude larger. However, those models seem to be ruled out by Planck already, due to their scalar spectral index alone, and therefore it is the remaining models which have such a small running.

In the not so far future, with larger range of scales coverage, e.g. with the measurement of the spectral index of linear fluctuations in neutral Hydrogen clouds emitting in the 21 cm line at and after reionization, by probes like the Square Kilometer Array~\cite{SKA}, we may be able to bring down the precision on $\alpha_s$ to the level of a few parts in 10,000, which could allow us to differentiate these models among themselves. Otherwise we will not have significant lever arm to tell them apart.

\section{Discussion}

In this paper we have introduced the $N$-formalism, which can be thought of as the background complement to the $\delta N$-formalism and where all inflationary observables are expressed in terms of the number of e-foldings. This is particularly useful given the generic requirement that $N$ should exceed $50$ and suggests a large-$N$ expansion. 

We have demonstrated that the majority of models  allow for such an expansion in $1/N$; the leading terms for $n_s$ are either constant, perturbative, non-perturbative or logarithmic. A quick overview of the properties of these classes can be found in table 1. Moreover, we have shown that subleading terms are generically irrelevant (in addition to ambiguously defined): the leading approximation at large-$N$ agrees excellently with the exact expression. The only exception to this behaviour are the non-perturbative examples of natural and quadratic hilltop. In these models, the effective expansion is in terms of $cN$, which does not allow for a leading approximation when $c$ is sufficiently small.

\begin{table}[ht]
\begin{center}
\begin{tabular}{||c||c||c||}
\hline 
Class & $V(\phi)$ & $\epsilon(N)$  \\ \hline
 \hline
Constant & $e^{\phi}$ & constant \\ \hline
Perturbative $p=1$ & $\phi^n$ with $n>0$ & $1/N$ \\ \hline
Perturbative $1<p<2$ & $1- \phi^n$ with $n<0$ & $1/N^p$ \\ \hline
Perturbative $p=2$ & $1-e^{-\phi}$ & $1/N^2$ \\ \hline
Perturbative $p>2$ & $1- \phi^n$ with $n>0$ & $1/N^p$ \\ \hline
Non-perturbative & $1-\phi^2$ & $e^{-N}$ \\ \hline
Logarithmic & $1-\phi e^{-\phi}$ & $\ln(N)/N$ \\ \hline
\end{tabular}
\end{center}
\caption{A sketch of the various classes and their corresponding asymptotic potentials and equations of state at large-$N$. More details can be found in section 3.}
\end{table}

We should also mention that there are inflationary models that do not allow for a large-$N$ expansion. The chief example is supergravity hybrid inflation~\cite{LindeRiotto} with a potential
\be
V(\phi) = \mu^4\left( 1 + {\kappa^2\over8\pi^2}\ln\frac{\phi}{\phi_c} + \frac{\phi^4}{8}\right) \,.
\ee
The resulting equation of state parameter reads
\be
\epsilon = \frac{\kappa^3}{16\pi^3\sin\frac{\kappa N}{\pi}(1+\cos\frac{\kappa N}{\pi})}\,.
\ee
Due to the trigonometric functions, this parameter does not have a definite large-$N$ behavior and hence defies the proposed classification. 

Building on the $N$-dependence of these universality classes, we have investigated the values of the running for the different inflationary models. Using the Planck constraints for the spectral index as input, it turns out that essentially all models give a similar prediction for the running of $n_s$ as a function of scale, centered around $\log_{10}|\alpha_s| = - 3.2$ and with a range of only half a decade. This remarkable feature seems to have gone unnoticed so far. As we have pointed out, one can gain a first understanding of the limited range of $\alpha_s$ from the preference of Planck for inflationary models with correct spectral indices for both $N=50$ as well as $N=60$. 

\section*{Note added}

A month after submission to the arXiv, the BICEP2 collaboration presented their ground-breaking results, and we had to adapt to the situation, changing slightly our perspective, without modifying our conclusions. In particular, a high-scale model of inflation makes the large-$N$ expansion more plausible, since a large reheating temperature imposes a robust lower limit on the number of e-folds. On the other hand, it shifts the attention from Starobinsky-like models to chaotic type models of inflation. But most importantly, it imposes very stringent constraints on a large array of models that fall under the various universality classes, as can be appreciated in the new Fig.~3.

\section*{Acknowledgements}

We thank Andrei Linde, Marco Scalisi and Ivonne Zavala for useful comments. We acknowledge financial support from the Madrid Regional Government (CAM) under the program HEPHACOS S2009/ESP-1473-02, from the Spanish MINECO under grant FPA2012-39684-C03-02 and Consolider-Ingenio 2010 PAU (CSD2007-00060), from the Centro de Excelencia Severo Ochoa Programme, under grant SEV-2012-0249, as well as from the European Union Marie Curie Initial Training Network UNILHC PITN-GA-2009-237920.

\appendix


\section{k-inflation class}

This class is characterized by both an equation of state parameter $\epsilon$ and a speed of sound $c_s^2=\epsilon\,\rho/(3X \rho_{,X})$ different from one, with $X=1/2(\partial\phi)^2$. The amplitude of the scalar power spectrum changes since it arises from a curvature fluctuation, $v=z\,\zeta$, satisfying a new mode equation
\be
v''_k + \Big(c_s^2 k^2 - \frac{z''}{z}\Big) v_k  = 0\,,
\ee
which gives
\be
P_s (k)  = \frac{H^2}{8\pi^2\,c_s \epsilon}\,,
\ee
while the tensor spectrum does not change. Therefore, the spectral indices, tensor ratio and scalar running are given by
\ba
n_s - 1 & = & -2\epsilon + \frac{(c_s\epsilon)'}{c_s\epsilon}\,,\\[2mm]
n_t & = & -2\epsilon \,,\\[2mm]
\frac{d \ln n_s}{d \ln k} & = & 2\epsilon' - \left(\frac{(c_s\epsilon)'}{c_s\epsilon}\right)'\,,\\[3mm]
r & = & 16\,c_s\epsilon = -8\,c_s\,n_g\,.
\ea

What is new in this model is a possibly large non-Gaussianity, $\zeta = \zeta_L - 3/5 f_{NL}\zeta_L^2$, which contributes mostly to the equilateral 3-point correlation function of fluctuations in the CMB. The present Planck data constrain $f_{NL}^{\rm eq}$ to satisfy
\be
f_{NL}^{\rm eq} = \frac{35}{108}\left(\frac{1}{c_s^2}-1\right) < 33  \hspace{5mm} \Rightarrow \hspace{5mm} c_s > 0.1\,.
\ee
We can combine these results with any of the previous models and find that at most we can reduce the tensor contribution one order of magnitude, but we should not expect a very big change in the value of the spectral indices.

\providecommand{\href}[2]{#2}\begingroup\raggedright\endgroup

\end{document}